\documentclass[aps,pre,twocolumn,superscriptaddress,showpacs,amsmath,amssymb]{revtex4-1}

\usepackage{graphicx}
\usepackage{color}

\begin{document}

\title{Fractality of eroded coastlines of correlated landscapes}

\author{P. A. Morais}
\email{pablo@fisica.ufc.br}
\affiliation{Departamento de F\'{\i}sica, Universidade Federal
  do Cear\'a, 60451-970 Fortaleza, Cear\'a, Brazil}

\author{E. A. Oliveira}
\email{erneson@fisica.ufc.br}
\affiliation{Departamento de F\'{\i}sica, Universidade Federal
  do Cear\'a, 60451-970 Fortaleza, Cear\'a, Brazil}

\author{N. A. M. Ara\'{u}jo}
\email{nuno@ethz.ch}
\affiliation{Computational Physics for Engineering Materials, IfB,
  ETH Z\"{u}rich, Schafmattstr. 6, 8093 Z\"{u}rich, Switzerland}

\author{H. J. Herrmann}
\email{hans@ifb.baug.ethz.ch}
\affiliation{Departamento de F\'{\i}sica, Universidade Federal
  do Cear\'a, 60451-970 Fortaleza, Cear\'a, Brazil}
\affiliation{Computational Physics for Engineering Materials, IfB,
  ETH Z\"{u}rich, Schafmattstr. 6, 8093 Z\"{u}rich, Switzerland}

\author{J. S. Andrade Jr.}
\email{soares@fisica.ufc.br}
\affiliation{Departamento de F\'{\i}sica, Universidade Federal
  do Cear\'a, 60451-970 Fortaleza, Cear\'a, Brazil}


\begin{abstract} 
  We investigate through numerical simulations of a simple sea-coast
  mechanical erosion model, the effect of spatial long-range
  correlations in the lithology of coastal landscapes on the fractal
  behavior of the corresponding coastlines. In the model, the
  resistance of a coast section to erosion depends on the local lithology
  configuration as well as on the number of neighboring sea sides. For
  weak sea forces, the sea is trapped by the coastline and the eroding
  process stops after some time. For strong sea forces erosion is
  perpetual. The transition between these two regimes takes place at a
  critical sea force, characterized by a fractal coastline front. For
  uncorrelated landscapes, we obtain, at the critical value,
  a fractal dimension $D=1.33$, which is consistent with the
  dimension of the accessible external perimeter of the spanning
  cluster in two-dimensional percolation. For sea forces above the
  critical value, our results indicate that the coastline is
  self-affine and belongs to the KPZ universality class. In the case
  of landscapes generated with power-law spatial long-range
  correlations, the coastline fractal dimension changes continuously
  with the Hurst exponent $H$, decreasing from $D=1.34$ to $1.04$, 
  for $H=0$ and $1$, respectively. This non-universal behavior is
  compatible with the multitude of fractal dimensions found for real
  coastlines.
\end{abstract}

\maketitle

\section{Introduction}

Since the introduction of the concept of fractals by Mandelbrot
\cite{Mandelbrot67}, scale invariant behavior has been identified and
investigated in many geological and geophysical
phenomena \cite{Turcotte89}, including the frequency distributions of
earthquakes \cite{Sornette94,*Saichev06,*Lippiello07,*Bottiglieri10}
and volcanic eruptions \cite{Kaminski98}, the size
distribution of rock fragments \cite{Astrom06,Kun06} and mineral deposits
\cite{Zhang01}, and the topography of river networks \cite{Banavar99,Dodds10}, rivers
deltas \cite{Seybold07,*Seybold10} and rocky coastlines \cite{Baldassarri08,Boffetta08}. More recently, the
 geometry of watersheds, namely, the lines separating
adjacent drainage basins (catchments), has been reported to display
also typical fractal features \cite{Fehr09,Fehr11,Andrade11}, with important
implications to water management \cite{Vorosmarty98,*Kwarteng00,*Sarangi05},
landslides \cite{Dhakal04,*Pradhan06,*Lazzari06,Lee06}, and flood prevention 
\cite{Burlando94,*Yang07}.

Erosion is certainly one of the most remarkable examples of a
geological process that naturally generates diverse self-similar structures.
In particular, the erosion of a coastline by the sea
\cite{Penning-Roswell92} constitutes a rather rich phenomenon.  Due
to the action of underlying geological processes (e.g., tectonics and
volcanic events), topological and lithological properties of coastal
landscapes are generally heterogeneous as well as long-range
correlated in space. As a consequence, the resistance to erosion must
be considered as a spatially dependent parameter. One should therefore
expect that the self-similar geometry of coastlines should emerge from
an intricate interplay between these landscape properties and the sea
force.

In a study by Sapoval {\it et al.} \cite{Sapoval04}, an erosion model
is proposed to show how surviving coastlines can dynamically evolve to
self-similar objects by means of a self-organized critical process
\cite{Bak96,Turcotte99}. By applying this model to spatially uncorrelated
landscapes, they recovered, at the critical steady-state, a fractal
dimension $D=1.33$ that is frequently observed in real systems
\cite{Boffetta08}. This dimension is different from $7/4$, the dimension
of the external perimeter of percolation \cite{Sapoval85}, but it is consistent
with $4/3$, the one for the accessible external perimeter \cite{Grossman87}.
This value, however, cannot be taken as universal, since, in fact, a multitude
of fractal dimensions has been measured for real coastlines
\cite{Richardson61}. 

In this work, we investigate the dependence of the fractal dimension
of sea-coast interfaces on the long-range correlations of
synthetic landscapes. In order to study the invasion of the sea
through the coast we consider a simple lattice invasion model. Each
coast site is characterized by a resistance to erosion which is a
function of its local lithology parameter and coastal configuration.
Spatial long-range correlated surfaces are generated with the Fourier
filtering method \cite{Peitgen88,Sahimi94,Sahimi96,Makse96,Oliveira11}, 
which allows to control the nature and the strength of correlations.

The manuscript is organized as follows. In Section~\ref{sec::model} we
introduce the model and its relevant definitions. Results for
correlated and uncorrelated distributions of the lithology are
discussed in Section~\ref{sec::results}. Finally, conclusions are draw
in Section~\ref{sec::conclusions}.

\section{The model}\label{sec::model}

  \begin{figure*}[t]
    \includegraphics[width=\textwidth]{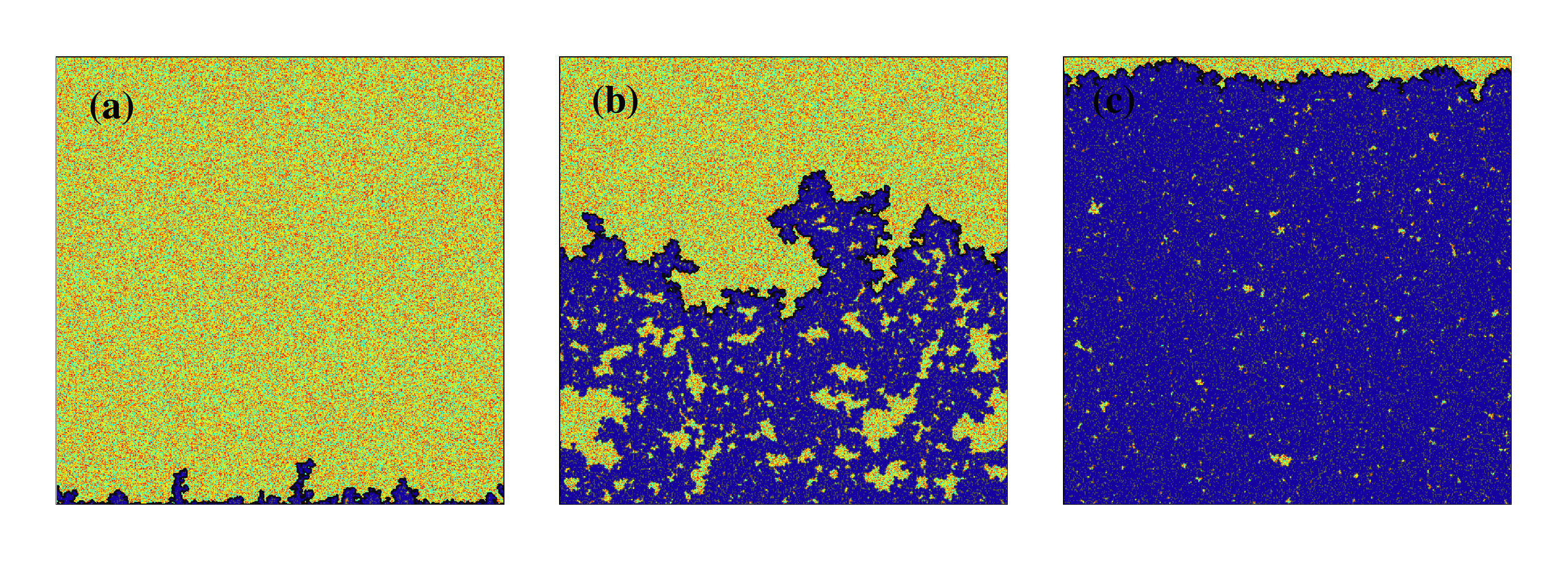}
    \caption{(color online) Snapshots of typical configurations
      obtained with the erosion model for uncorrelated lithology on a
      lattice with $512^2$ sites. The sea sites are in blue (dark
      grey) and all the other sites are land sites (light grey).
      Three different regimes are obtained: a) subcritical, weak sea
      force, regime ($f<f_c$): the coastline is rough but not fractal.
      b) critical regime ($f=f_c$): the coastline is a self-similar
      fractal.  c) supercritical, strong sea force, regime ($f>f_c$):
      the coastline is no longer self-similar, but rather self-affine.
      \label{fig::snapshots}
    }
  \end{figure*}

\begin{figure}[b]
  \includegraphics[width=\columnwidth]{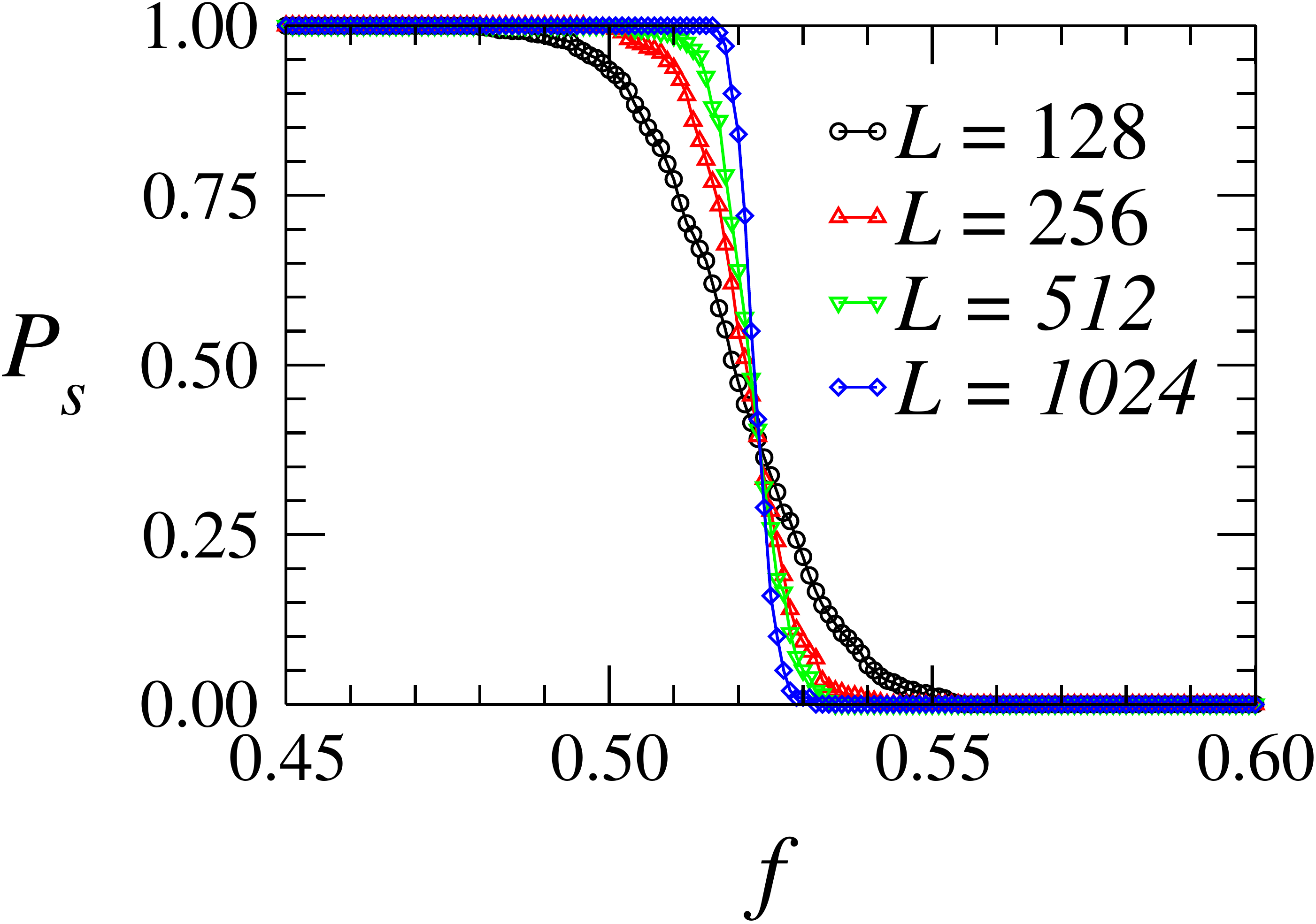}
  \caption{(color online) Dependence of the sea-trapping probability,
    $P_s$, on the erosion force $f$, for landscapes generated with a uniform
    distribution of the lithology parameter $\ell$. A transition is observed
    from a trapped, subcritical, regime -- for weak sea force -- to a
    perpetual, supercritical, invasion regime -- for strong sea force.
    The transition occurs for $f_c=0.523\pm0.001$ -- estimated
    from the crossing of the lines.  Each curve corresponds to a
    different system size $L^2$, with $L=\{128,256,512,1024\}$, and
    results have been averaged over $\{800,400,200,100\}$ samples,
    respectively.
    \label{fig::traprob}
  }
\end{figure}

\begin{figure}[b]
  \includegraphics[width=\columnwidth]{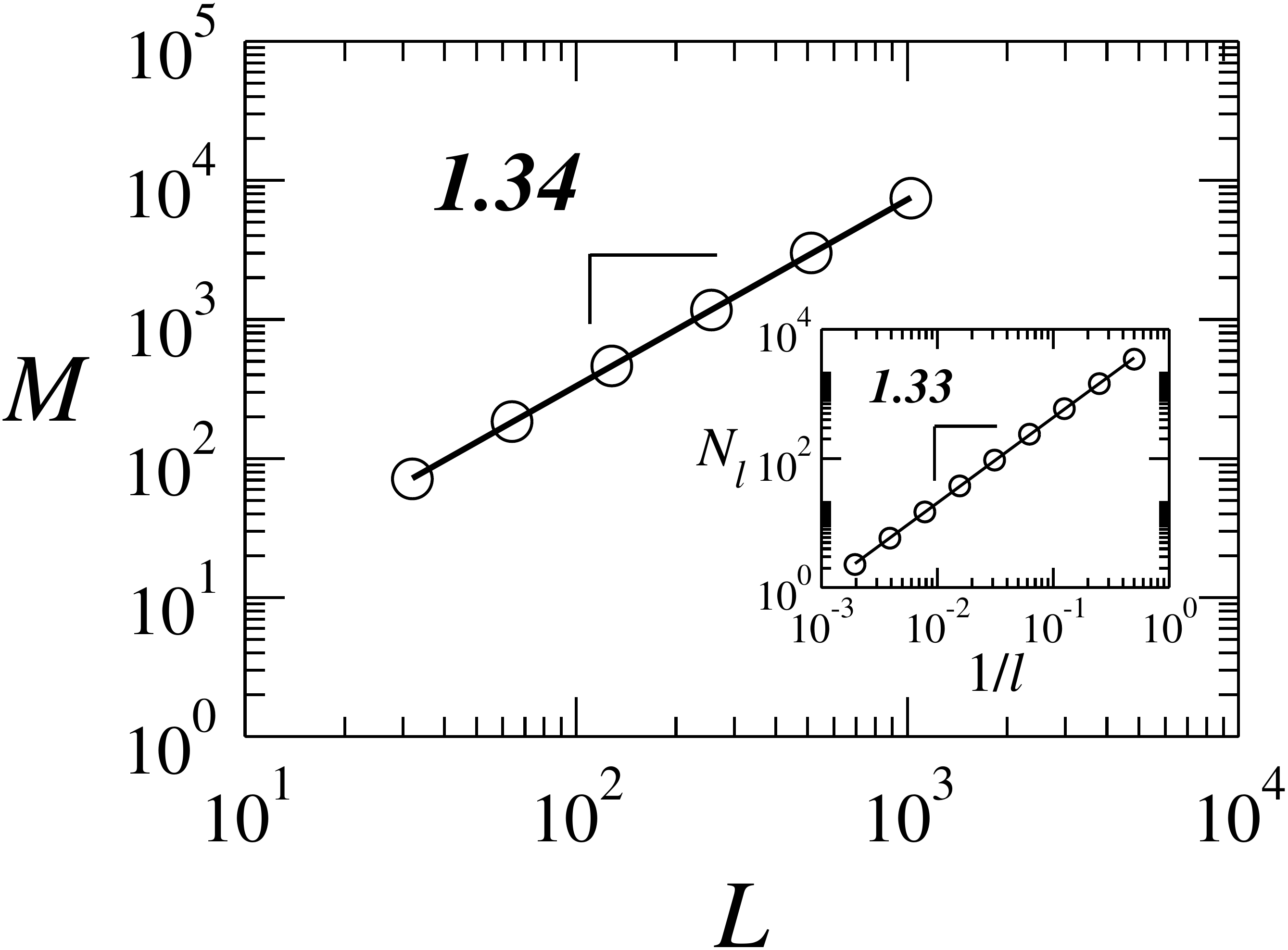}
  \caption{Size dependence of the mass of the coastline -- number of
    sites -- at the critical sea force, $f=f_c$. A fractal dimension
    $D=1.34\pm0.01$ is obtained from the best fit of the data
    points.  Square lattices, with uniform distribution of the
    lithology parameter, of size $L^2$ have been considered with
    $L=\{32,64,128,256,512,1024\}$. Results have been averaged over
    $\{3200,1600,800,400,200,100\}$ independent realizations.  The
    inset shows, for the yardstick method, the number of sticks needed
    to cover the line as a function of the inverse of the stick size,
    for a system with $L=1024$, averaged over $10^2$ samples.
    From the fit, a fractal dimension $D=1.33\pm0.01$ is obtained.
    From both methods the fractal dimension of the coastline in the
    critical regime is estimated to be $D=1.33\pm0.01$.
    \label{fig::fractaldim}}
\end{figure}

In a recent
work, Sapoval \textit{et al.} \cite{Sapoval04} proposed a model that
explicitly split the dynamics of erosion into two different mechanism
based on their characteristic time scale, namely, slow and rapid
dynamics.  While the former is mainly related to chemical processes,
the latter is solely due to mechanical erosion. Here we only focus on
the rapid dynamics, since the slow one takes place on a time scale
which is beyond the scope of this study.

For simplicity, the system is mapped onto a regular square lattice,
where each site can be either a sea or a land site.  For land coast
site, a lithology parameter $\ell_i$ is assigned, which coarse-grains
several geological properties, characterizing the mechanical
interaction with the sea.  We assume that an island site completely
surrounded by the sea is more fragile than a similar one in the
coastline, so that the resistance to erosion $r_i$ depends on the local
coastal configuration as,
\begin{equation}
\label{eq::resistance}
r_i=\ell_i^{n_i} \ \ ,
\end{equation}
where $n_i$ is the number of neighboring sea sites. Accordingly, a
coast site with only one sea neighbor has resistance $r_i=\ell_i$, while
one completely clipped by water has resistance $r_i=\ell_i^4$. During the
invasion process, the sea penetrates in the coastline with a constant
force $f$. Initially, the coastline, defined as the interface between
sea and land, is a straight line at the bottom of the system and
periodic boundary conditions are applied in the horizontal direction.
At each iteration, all coast sites in the neighborhood of a sea site,
and with a resistance to erosion below the sea force, become part of
the sea.
The proposed model corresponds to a collective invasion percolation where 
the resistance is a function of the number of neighboring sea sites.
For $r_i=l_i$ the resistance is solely dependent on the lithology
parameter and ordinary percolation is recovered.
The introduced weakening mechanism promotes the erosion of earth filaments
in the coast, typically observed for ordinary percolation.

\section{Results}\label{sec::results}

\begin{figure}
  \includegraphics[width=\columnwidth]{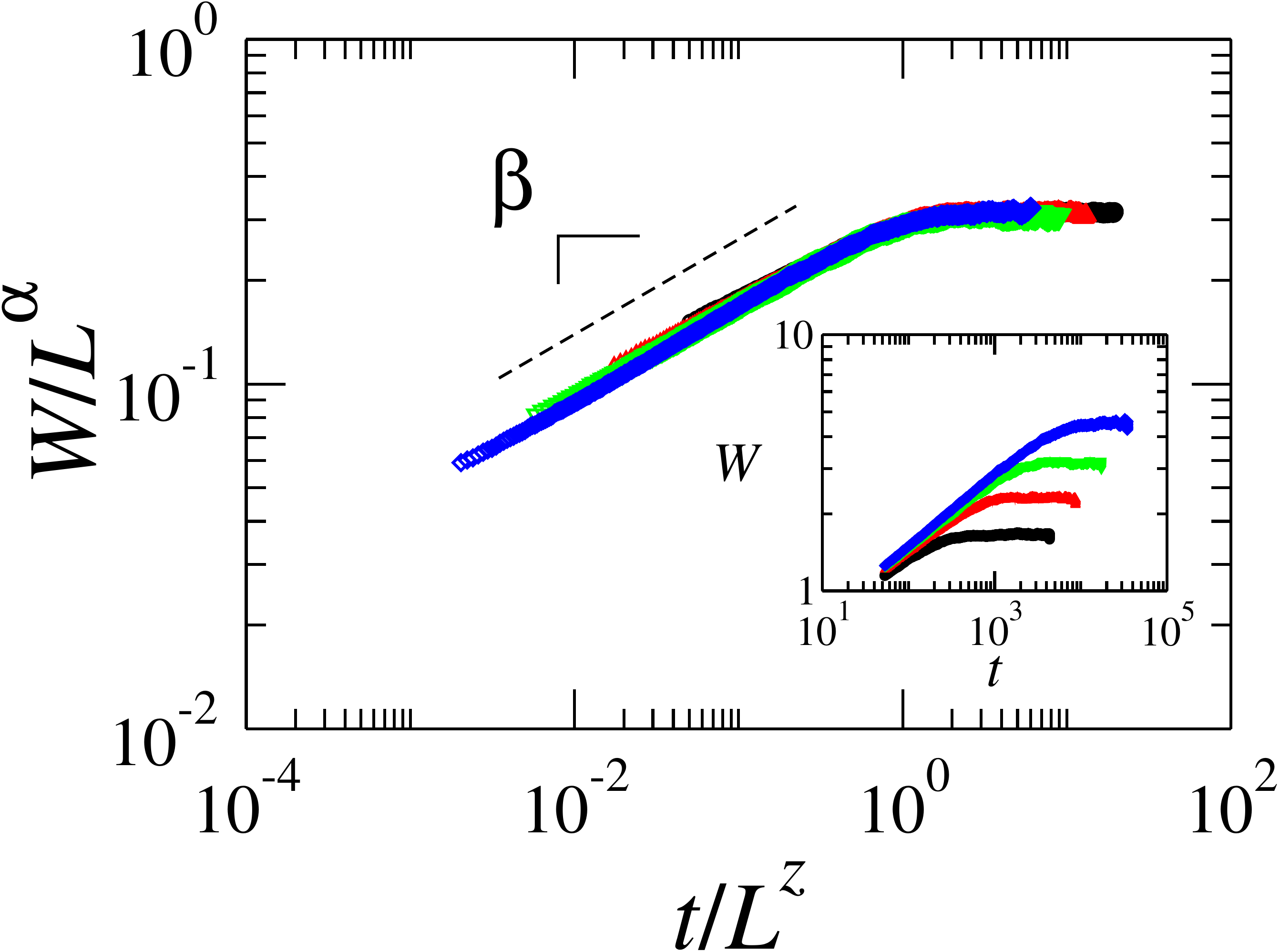}
  \caption{(color online) Finite-size scaling of the time dependence of the
    coastline roughness in the supercritical regime ($f=0.9$).  
    Square lattices  with $L_x \times L_y$ sites have been considered, 
    with $L_x=L$ and $L_y=128L$. The time $t$ has been rescaled by $L^z$,
    where $z$ is the dynamic exponent, and the roughness has been
    rescaled in units of $L^\alpha$, where $\alpha$ is the roughness
    exponent. The best data collapse is obtained for
    $\alpha=0.48\pm0.04$ and $z=1.55\pm0.11$, a value that is consistent with the
    Kardar-Parisi-Zhang universality class \cite{Kardar86}.  
    The straight line has slope $\beta$, corresponding to the growth exponent.  
    The inset shows the time dependence of the coastline roughness.  Systems
    with $L=\{32,64,128,256\}$ have been sampled over
    $\{3200,1600,800,400\}$ independent runs, respectively.
    \label{fig::sizescaling}}
\end{figure}

\begin{figure}
  \includegraphics[width=\columnwidth]{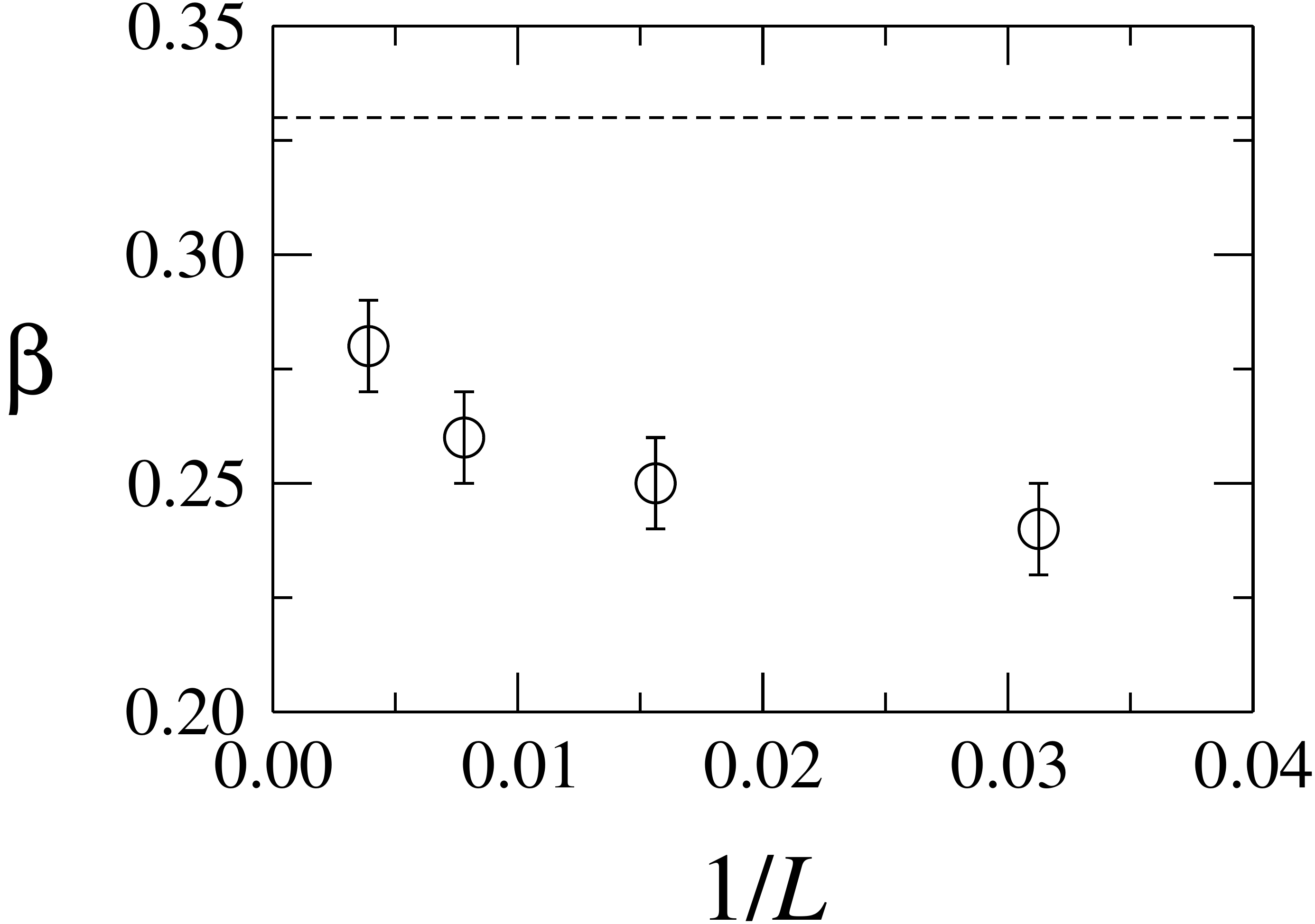}
  \caption{Size dependence of the growth exponent, $\beta$.  These results
    indicate that, in the thermodynamic limit, $\beta$ converges to $\beta=1/3$
    (dashed line) as in the Kardar-Parisi-Zhang universality class. Systems with
    $L=\{32,64,128,256\}$ have been sampled over
    $\{3200,1600,800,400\}$ independent runs,
    respectively.\label{fig::growthexp}}
\end{figure}

\begin{figure*}
  \includegraphics[width=0.9\textwidth]{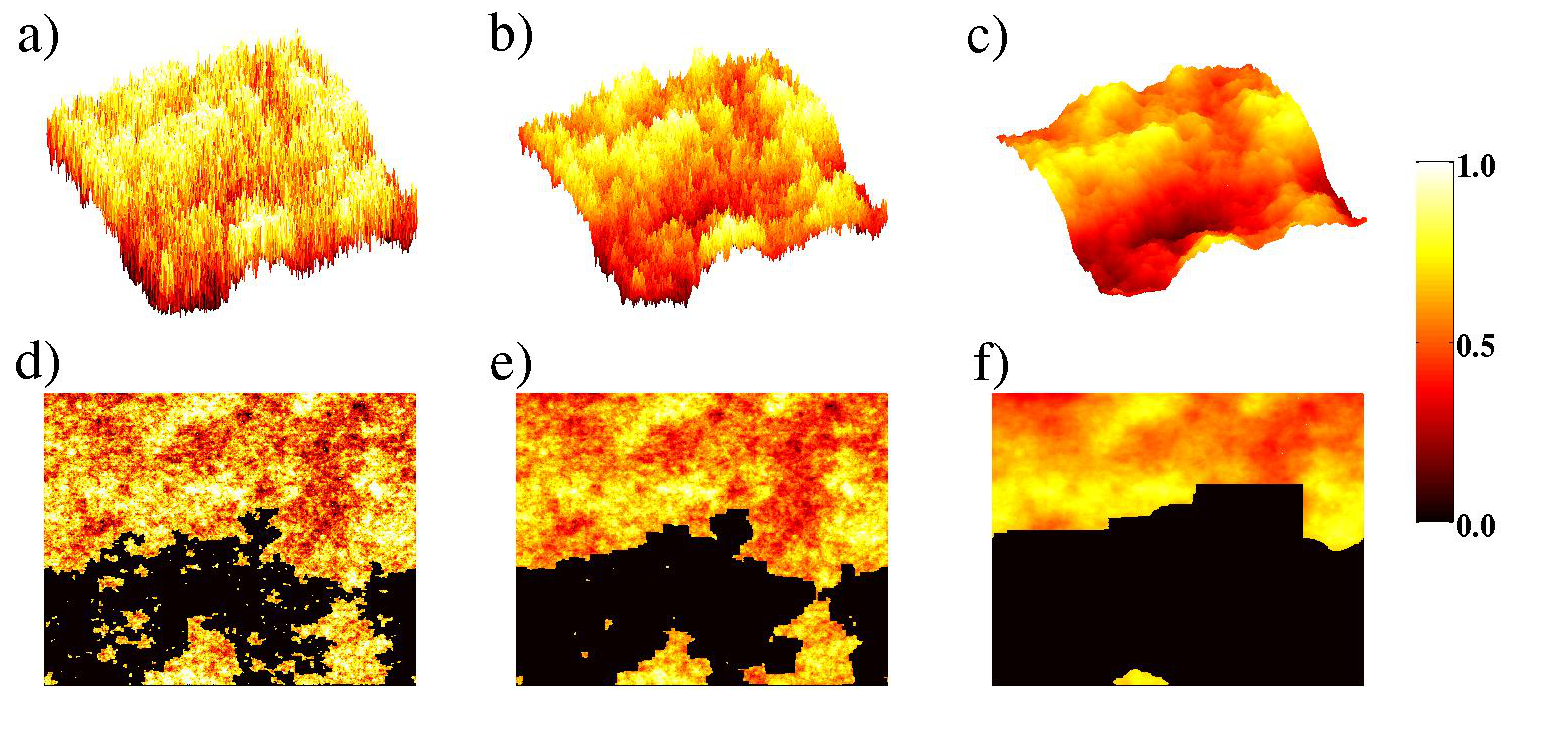}
  \caption{(color online) Snapshots of the initial distribution of lithology 
    parameter (a, b, and c) and the corresponding eroded system (d, e, and f) 
    at the critical force ($f=f_c$), for three landscapes generated with 
    long-range spatial correlations, for different Hurst exponent: 
    a) and d) $H=0.0$; b) and e) $H=0.2$; and c) and f) $H=0.8$.
    The color scheme of coast sites represents the value of the lithology 
    parameter $l$ and sea sites were assigned with $l=0$. Pictures have been
    obtained for square lattices with $512^2$ sites.
    \label{fig::snapcorr}}
\end{figure*}

  \subsection{Uncorrelated lithology}

Let us start with the study of the model on an uncorrelated lithology,
where the lithology parameter is uniformly distributed in the range
$0<\ell<1$. 
Figure~\ref{fig::snapshots} shows a typical configuration of the
system for different sea forces $f$ and obtained for the same
distribution of the lithology parameter. Two different regimes are
observed based on the sea force, namely, a weak and a strong regime.
For weak sea force, as shown in Fig.~\ref{fig::snapshots}(a),
corresponding to low values of $f$, only few sites are eroded and
invaded by the sea.  After some finite number of steps, the sea is
completely trapped by the coastline -- the resistance of all coastline
sites is greater than the sea force -- and the sea cannot invade
further. For strong sea force, Fig.~\ref{fig::snapshots}(c), the
coastline perpetually grows, without being ever trapped.
Nevertheless, some islands remain that will never be destroyed, i.e.,
their coastline sites have a resistance above the sea force. 
As discussed before, in reality two different mechanisms occur: the mechanical
and the chemical erosion. In this work, we solely account for the former one,
however, over larger time scales, due to the latter mechanism the sea is
permanently eroding the coastline.

The transition between the weak and the strong sea force regimes
occurs at a critical force $f=f_c$. To determine this critical force,
we analyze the sea-trapping probability defined as the probability,
for a given sea force, for an erosion process to be limited to a finite number of
iterations. In Fig.~\ref{fig::traprob} we show this probability as a
function of the sea force, for different system sizes. For the weak sea
force, $f<f_c$, since the sea invasion is always halted after some
steps, the probability is one. On the other hand, in the strong sea force
regime, $f>f_c$, the sea erosion is never stopped.  From the crossing
of all lines (different system sizes), it is possible to estimate the
critical force as $f_c=0.523\pm0.001$.  
At the critical force, the front of the coastline is self-similar with
a fractal dimension $D=1.33\pm0.01$. To obtain the fractal
dimension, we considered two different methods: the scaling
of the set of coastal sites with system size (main plot of
Fig.~\ref{fig::fractaldim}) and the yardstick method (inset of
Fig.~\ref{fig::fractaldim}) \cite{Tricot88}.
The critical force obtained under the proposed weakening mechanism is
lower than the one corresponding to ordinary percolation ($f_c\approx0.593$),
recovered when the resistance is solely dependent on the lithology parameter, in the absence of weakening.
In such case, the obtained fractal dimension is compatible with $7/4$, the one of
the hull of the percolation cluster \cite{Grossman87}.

For strong sea force, $f>f_c$, the sea constantly erodes the coast,
being never trapped, and the resulting coastline front is self-affine.
Moreover, our results indicate that the statistical properties of the
sea invasion process belong to the same universality class of
interfaces that obey the Kardar-Parisi-Zhang (KPZ) equation
\cite{Kardar86}. To characterize the main interface (coastline
front), we study the time evolution of the roughness, $W$, namely, the
standard deviation of the interface width, defined as,
\begin{equation}
W=\sqrt{\frac{1}{L}\sum_{i=1}^L\left(y_i-\langle y\rangle\right)^2} \ \ ,
\end{equation}
\noindent where $y_i$ is the vertical position of the interface and
$\langle y\rangle$ its mean value over all columns.
Results for different system sizes are shown in the inset of
Fig.~\ref{fig::sizescaling}. Initially, the roughness is an algebraic
function of time, $W\sim t^\beta$, where $\beta$ is the growth
exponent. At a certain crossover, $t=t_\times$, lateral correlations
in space resulting from the invasion process become so large, as
compared to the finite size $L$, that the roughness reaches a
saturation value, $W_\mathrm{sat}$ \cite{Barabasi95}.
With the system size, $W_\mathrm{sat}$ diverges as $W_\mathrm{sat}\sim
L^\alpha$ and the crossover time scales as $t_\times\sim L^z$, where
$\alpha$ is the roughness exponent and $z$ the dynamic exponent.
Therefore, a complete finite-size scaling can be obtained with the
\textit{ansatz} \cite{Family85},
\begin{equation}\label{eq::scaling}
W(L,t)=L^\alpha\left(\frac{t}{L^z}\right) \ \ .
\end{equation}
The main plot of Fig.~\ref{fig::sizescaling} corresponds to the
finite-size scaling of the roughness. From the data collapse,
we obtain the values $\alpha=0.48\pm0.04$ and $z=1.55\pm0.11$, which
are consistent with the exponents $\alpha=1/2$ and $z=3/2$ from the
KPZ universality class \cite{Kardar86,Odor04}.  The size dependence of
the growth exponent is shown in Fig.~\ref{fig::growthexp}.  Our results indicate
that, in the thermodynamic limit, the growth exponent converges to the
expected KPZ value of $\beta=1/3$.
For strong sea force, when the weakening by neighboring sea
sites is neglected, the supercritical regime of ordinary percolation
is recovered.

\subsection{Correlated lithology}

\begin{figure}[t]
  \includegraphics[width=\columnwidth]{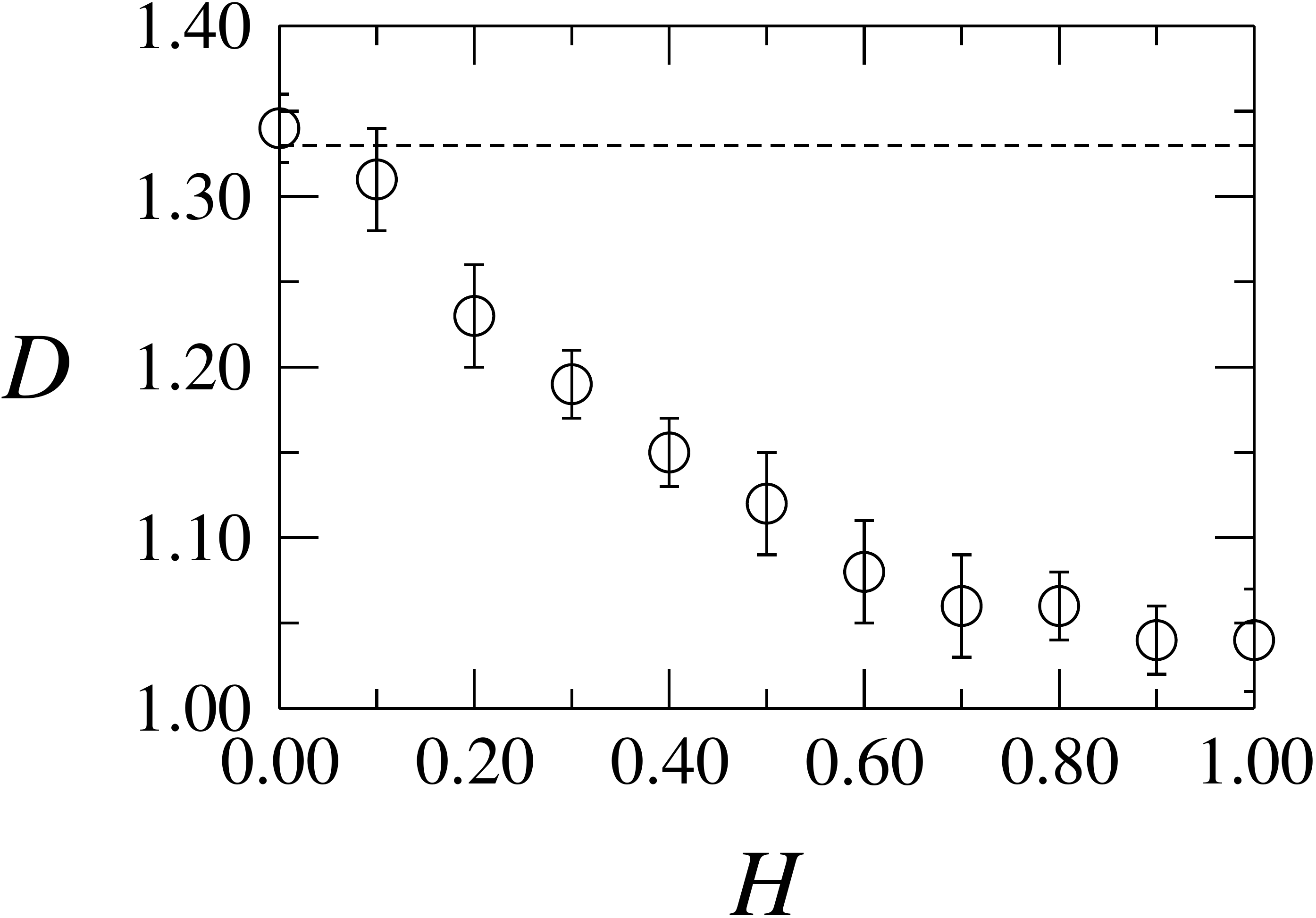}
  \caption{
    Critical coastline fractal dimension $D$ as a function of
    the Hurst exponent $H$.  A continuous decrease from $D=1.34$ at $H=0$
    to $D=1.04$ at $H=1.0$ is observed.
    The dashed line stands for the fractal dimension of the
    uncorrelated case ($D=1.33$).  Each point is obtained from the
    size dependence of the number of coastline sites for system sizes
    $L=\{32,64,128,256,512,1024\}$ averaged over 
    $\{3200,1600,800,400,200,100\}$ samples.
    \label{fig::fractaldimhurst}}
\end{figure}

In this section we discuss the case where the disordered coastal
landscapes possess spatial long-range correlations in the lithology
parameter $\ell$.  As in previous studies
\cite{Prakash92,Sahimi94,Sahimi96,Makse96,Kikkinides99,Stanley99,Makse00,Araujo02,Araujo03,Du04,Oliveira11},
spatial long-range correlated distributions are obtained with
fractional Brownian motion (fBm) \cite{Peitgen88,Mandelbrot67}.  To
achieve the intended correlated distribution, Fourier coefficients are
generated in the reciprocal space of frequencies $f$, according to a
power-law spectral density, namely,
\begin{equation}
S(f_1,\cdots,f_d)=\left(\sqrt{\sum_{i=1}^{d}f_i^2}\right)^{-w} \ \ ,
\end{equation}
\noindent where $d$ is the spatial dimension ($d=2$ in this work),
 and the inverse Fourier transform is applied to obtain the
distribution in real space.  Several samples are then generated
and the distributions truncated between $-3\sigma$ and $3\sigma$,
where $\sigma$ is the standard deviation.  The truncation is such that
values outside this range are assigned to be $\pm3\sigma$, keeping the
original sign.  Finally, the distribution is rescaled in the interval
$[0\!:\!1]$. Each distribution is characterized by a Hurst
exponent $H$ -- related to the spectral exponent by $w=2H+d$ -- 
such that, for two dimensions, the
correlations are negative for $0<H<1/2$ and positive for $1/2<H<1$.
The former case means that neighbors of a strong site are, on average,
weak sites, whereas in the latter case they are typically strong sites.
For $H=1/2$, the classical Brownian motion is recovered where the increments
are uncorrelated but the obtained lithology is still correlated.
The uncorrelated distribution of lithology is solely obtained for a
constant spectral density, with $w=0$ and $H=-d/2$ ($H=-1$ in two 
dimensions). Further details about the adopted methodology can be 
found, for example, in Ref.~\cite{Oliveira11}.

Figure~\ref{fig::snapcorr} shows snapshots of the system, at the
critical force, for three different values of the Hurst exponent,
namely, $0$, $0.2$, and $0.8$. In the first two the spatial
correlations are negative, while the latter corresponds to a strong
positive correlation. By increasing the Hurst exponent, islands
become less frequent and the interface smooths out.
In the limit of very strong positive correlations, the coastline 
converges to a non-fractal object with dimension one. In
Fig.~\ref{fig::fractaldimhurst} we show the fractal dimension of the
coastline front $D$ as a function of the Hurst exponent.  The
fractal dimension decreases continuously from $D=1.34\pm0.02$, for
$H=0$, to $D=1.04\pm0.03$, for strong positive correlations.

\section{Conclusions}\label{sec::conclusions}

In this work, a model is introduced to study the mechanical erosion of
the coastline by the sea.  Despite sharing some features with 
invasion percolation, in this model the update occurs over the entire
interface and not only at a single interface site and the resistance 
of a coast site to erosion depends on the local
configuration. The larger the number of neighboring sea sites the
weaker the resistance.  We have shown that, based on the sea force,
the system can either be in the weak or strong sea force regime.
While in the former the sea is trapped by the coastline and the
eroding process stops after some time, in the latter, erosion is
perpetual and belongs to the Kardar-Parisi-Zhang universality class.
The transition between these two regimes occurs at a critical sea
force, characterized by a fractal coastline front. For an uncorrelated
distribution of the coast lithological properties, the fractal
dimension of this interface is related with the accessible perimeter
of the percolation cluster, whereas for coasts with long-range
correlation, the fractal dimension changes with the Hurst exponent.
In Ref.~\cite{Sapoval04} a model with a self-stabilization mechanism is
proposed to explain how the system is found at criticality. Here,
we solely focus on the effect of correlations on the properties of the 
critical state without self-organization. Yet, this study clarifies the 
relation between the spatial correlations in the lithology and the multitude
of fractal dimensions observed for coastlines. 
As an extension of this work, one might consider the effect of disorder in
the strong sea force regime.
\begin{acknowledgments}
  We thank Bernard Sapoval and Andrea Baldassarri for fruitful
  discussions. We also thank the ETH Competence Center Coping with
  Crises in Complex Socio-Economic Systems (CCSS) through the ETH
  Research Grant CH1-01-08-2, the Brazilian agencies CNPq, CAPES and
  FUNCAP, and the Pronex grant CNPq/FUNCAP, for financial support.
\end{acknowledgments}

\bibliography{erosion}

%merlin.mbs 2010-03-15 4.21a (PWD, AO, DPC)
\begin{thebibliography}{52}%
\makeatletter
\providecommand \@ifxundefined [1]{%
 \@ifx{#1\undefined}
}%
\providecommand \@ifnum [1]{%
 \ifnum #1\expandafter \@firstoftwo
 \else \expandafter \@secondoftwo
 \fi
}%
\providecommand \@ifx [1]{%
 \ifx #1\expandafter \@firstoftwo
 \else \expandafter \@secondoftwo
 \fi
}%
\providecommand \natexlab [1]{#1}%
\providecommand \enquote  [1]{``#1''}%
\providecommand \bibnamefont  [1]{#1}%
\providecommand \bibfnamefont [1]{#1}%
\providecommand \citenamefont [1]{#1}%
\providecommand \href@noop [0]{\@secondoftwo}%
\providecommand \href [0]{\begingroup \@sanitize@url \@href}%
\providecommand \@href[1]{\@@startlink{#1}\@@href}%
\providecommand \@@href[1]{\endgroup#1\@@endlink}%
\providecommand \@sanitize@url [0]{\catcode `\\12\catcode `\$12\catcode
  `\&12\catcode `\#12\catcode `\^12\catcode `\_12\catcode `\%12\relax}%
\providecommand \@@startlink[1]{}%
\providecommand \@@endlink[0]{}%
\providecommand \url  [0]{\begingroup\@sanitize@url \@url }%
\providecommand \@url [1]{\endgroup\@href {#1}{\urlprefix }}%
\providecommand \urlprefix  [0]{URL }%
\providecommand \Eprint [0]{\href }%
\@ifxundefined \urlstyle {%
  \providecommand \doi  [0]{\begingroup \@sanitize@url \@doi}%
  \providecommand \@doi [1]{\endgroup \@@startlink {\doibase
  #1}doi:\discretionary {}{}{}#1\@@endlink }%
}{%
  \providecommand \doi  [0]{doi:\discretionary{}{}{}\begingroup
  \urlstyle{rm}\Url }%
}%
\providecommand \doibase [0]{http://dx.doi.org/}%
\providecommand \Doi [0]{\begingroup \@sanitize@url \@Doi }%
\providecommand \@Doi  [1]{\endgroup\@@startlink{\doibase#1}\@@Doi}%
\providecommand \@@Doi [1]{#1\@@endlink}%
\providecommand \selectlanguage [0]{\@gobble}%
\providecommand \bibinfo  [0]{\@secondoftwo}%
\providecommand \bibfield  [0]{\@secondoftwo}%
\providecommand \translation [1]{[#1]}%
\providecommand \BibitemOpen [0]{}%
\providecommand \bibitemStop [0]{}%
\providecommand \bibitemNoStop [0]{.\EOS\space}%
\providecommand \EOS [0]{\spacefactor3000\relax}%
\providecommand \BibitemShut  [1]{\csname bibitem#1\endcsname}%
%</preamble>
\bibitem [{\citenamefont {Mandelbrot}(1967)}]{Mandelbrot67}%
  \BibitemOpen
  \bibfield  {author} {\bibinfo {author} {\bibfnamefont {B.~B.}\ \bibnamefont
  {Mandelbrot}},\ }\href@noop {} {\bibfield  {journal} {\bibinfo  {journal}
  {Science},\ }\textbf {\bibinfo {volume} {156}},\ \bibinfo {pages} {636}
  (\bibinfo {year} {1967})}\BibitemShut {NoStop}%
\bibitem [{\citenamefont {Turcotte}(1989)}]{Turcotte89}%
  \BibitemOpen
  \bibfield  {author} {\bibinfo {author} {\bibfnamefont {D.~L.}\ \bibnamefont
  {Turcotte}},\ }\href@noop {} {\bibfield  {journal} {\bibinfo  {journal} {Pure
  Appl. Geophys.},\ }\textbf {\bibinfo {volume} {131}},\ \bibinfo {pages} {171}
  (\bibinfo {year} {1989})}\BibitemShut {NoStop}%
\bibitem [{\citenamefont {Sornette}(1994)}]{Sornette94}%
  \BibitemOpen
  \bibfield  {author} {\bibinfo {author} {\bibfnamefont {D.}~\bibnamefont
  {Sornette}},\ }\href@noop {} {\bibfield  {journal} {\bibinfo  {journal}
  {Phys. Rev. Lett.},\ }\textbf {\bibinfo {volume} {72}},\ \bibinfo {pages}
  {2306} (\bibinfo {year} {1994})}\BibitemShut {NoStop}%
\bibitem [{\citenamefont {Saichev}\ and\ \citenamefont
  {Sornette}(2006)}]{Saichev06}%
  \BibitemOpen
  \bibfield  {author} {\bibinfo {author} {\bibfnamefont {A.}~\bibnamefont
  {Saichev}}\ and\ \bibinfo {author} {\bibfnamefont {D.}~\bibnamefont
  {Sornette}},\ }\href@noop {} {\bibfield  {journal} {\bibinfo  {journal}
  {Phys. Rev. Lett.},\ }\textbf {\bibinfo {volume} {97}},\ \bibinfo {pages}
  {078501} (\bibinfo {year} {2006})}\BibitemShut {NoStop}%
\bibitem [{\citenamefont {Lippiello}\ \emph {et~al.}(2007)\citenamefont
  {Lippiello}, \citenamefont {Godano},\ and\ \citenamefont {\mbox{de
  Arcangelis}}}]{Lippiello07}%
  \BibitemOpen
  \bibfield  {author} {\bibinfo {author} {\bibfnamefont {E.}~\bibnamefont
  {Lippiello}}, \bibinfo {author} {\bibfnamefont {C.}~\bibnamefont {Godano}}, \
  and\ \bibinfo {author} {\bibfnamefont {L.}~\bibnamefont {\mbox{de
  Arcangelis}}},\ }\href@noop {} {\bibfield  {journal} {\bibinfo  {journal}
  {Phys. Rev. Lett.},\ }\textbf {\bibinfo {volume} {98}},\ \bibinfo {pages}
  {098501} (\bibinfo {year} {2007})}\BibitemShut {NoStop}%
\bibitem [{\citenamefont {Bottiglieri}\ \emph {et~al.}(2010)\citenamefont
  {Bottiglieri}, \citenamefont {\mbox{de Arcangelis}}, \citenamefont {Godano},\
  and\ \citenamefont {Lippiello}}]{Bottiglieri10}%
  \BibitemOpen
  \bibfield  {author} {\bibinfo {author} {\bibfnamefont {M.}~\bibnamefont
  {Bottiglieri}}, \bibinfo {author} {\bibfnamefont {L.}~\bibnamefont {\mbox{de
  Arcangelis}}}, \bibinfo {author} {\bibfnamefont {C.}~\bibnamefont {Godano}},
  \ and\ \bibinfo {author} {\bibfnamefont {E.}~\bibnamefont {Lippiello}},\
  }\href@noop {} {\bibfield  {journal} {\bibinfo  {journal} {Phys. Rev.
  Lett.},\ }\textbf {\bibinfo {volume} {104}},\ \bibinfo {pages} {158501}
  (\bibinfo {year} {2010})}\BibitemShut {NoStop}%
\bibitem [{\citenamefont {Kaminski}\ and\ \citenamefont
  {Jaupart}(1998)}]{Kaminski98}%
  \BibitemOpen
  \bibfield  {author} {\bibinfo {author} {\bibfnamefont {E.}~\bibnamefont
  {Kaminski}}\ and\ \bibinfo {author} {\bibfnamefont {C.}~\bibnamefont
  {Jaupart}},\ }\href@noop {} {\bibfield  {journal} {\bibinfo  {journal} {J.
  Geophys. Res.},\ }\textbf {\bibinfo {volume} {103}},\ \bibinfo {pages}
  {29759} (\bibinfo {year} {1998})}\BibitemShut {NoStop}%
\bibitem [{\citenamefont {Astr\"om}(2006)}]{Astrom06}%
  \BibitemOpen
  \bibfield  {author} {\bibinfo {author} {\bibfnamefont {J.~A.}\ \bibnamefont
  {Astr\"om}},\ }\href@noop {} {\bibfield  {journal} {\bibinfo  {journal} {Adv.
  Phys.},\ }\textbf {\bibinfo {volume} {55}},\ \bibinfo {pages} {247} (\bibinfo
  {year} {2006})}\BibitemShut {NoStop}%
\bibitem [{\citenamefont {Kun}\ \emph {et~al.}(2006)\citenamefont {Kun},
  \citenamefont {Wittel}, \citenamefont {Herrmann}, \citenamefont {Kr\"oplin},\
  and\ \citenamefont {Mal\o{}y}}]{Kun06}%
  \BibitemOpen
  \bibfield  {author} {\bibinfo {author} {\bibfnamefont {F.}~\bibnamefont
  {Kun}}, \bibinfo {author} {\bibfnamefont {F.~K.}\ \bibnamefont {Wittel}},
  \bibinfo {author} {\bibfnamefont {H.~J.}\ \bibnamefont {Herrmann}}, \bibinfo
  {author} {\bibfnamefont {B.~H.}\ \bibnamefont {Kr\"oplin}}, \ and\ \bibinfo
  {author} {\bibfnamefont {K.~J.}\ \bibnamefont {Mal\o{}y}},\ }\href@noop {}
  {\bibfield  {journal} {\bibinfo  {journal} {Phys. Rev. Lett.},\ }\textbf
  {\bibinfo {volume} {96}},\ \bibinfo {pages} {025504} (\bibinfo {year}
  {2006})}\BibitemShut {NoStop}%
\bibitem [{\citenamefont {Zhang}\ \emph {et~al.}(2001)\citenamefont {Zhang},
  \citenamefont {Mao},\ and\ \citenamefont {Cheng}}]{Zhang01}%
  \BibitemOpen
  \bibfield  {author} {\bibinfo {author} {\bibfnamefont {Z.~R.}\ \bibnamefont
  {Zhang}}, \bibinfo {author} {\bibfnamefont {H.~H.}\ \bibnamefont {Mao}}, \
  and\ \bibinfo {author} {\bibfnamefont {Q.~M.}\ \bibnamefont {Cheng}},\
  }\href@noop {} {\bibfield  {journal} {\bibinfo  {journal} {Math. Geol.},\
  }\textbf {\bibinfo {volume} {33}},\ \bibinfo {pages} {217} (\bibinfo {year}
  {2001})}\BibitemShut {NoStop}%
\bibitem [{\citenamefont {Banavar}\ \emph {et~al.}(1999)\citenamefont
  {Banavar}, \citenamefont {Maritan},\ and\ \citenamefont
  {Rinaldo}}]{Banavar99}%
  \BibitemOpen
  \bibfield  {author} {\bibinfo {author} {\bibfnamefont {J.~R.}\ \bibnamefont
  {Banavar}}, \bibinfo {author} {\bibfnamefont {A.}~\bibnamefont {Maritan}}, \
  and\ \bibinfo {author} {\bibfnamefont {A.}~\bibnamefont {Rinaldo}},\
  }\href@noop {} {\bibfield  {journal} {\bibinfo  {journal} {Nature},\ }\textbf
  {\bibinfo {volume} {399}},\ \bibinfo {pages} {130} (\bibinfo {year}
  {1999})}\BibitemShut {NoStop}%
\bibitem [{\citenamefont {Dodds}(2010)}]{Dodds10}%
  \BibitemOpen
  \bibfield  {author} {\bibinfo {author} {\bibfnamefont {P.~S.}\ \bibnamefont
  {Dodds}},\ }\href@noop {} {\bibfield  {journal} {\bibinfo  {journal} {Phys.
  Rev. Lett.},\ }\textbf {\bibinfo {volume} {104}},\ \bibinfo {pages} {048702}
  (\bibinfo {year} {2010})}\BibitemShut {NoStop}%
\bibitem [{\citenamefont {Seybold}\ \emph {et~al.}(2007)\citenamefont
  {Seybold}, \citenamefont {\mbox{Andrade Jr.}},\ and\ \citenamefont
  {Herrmann}}]{Seybold07}%
  \BibitemOpen
  \bibfield  {author} {\bibinfo {author} {\bibfnamefont {H.}~\bibnamefont
  {Seybold}}, \bibinfo {author} {\bibfnamefont {J.~S.}\ \bibnamefont
  {\mbox{Andrade Jr.}}}, \ and\ \bibinfo {author} {\bibfnamefont {H.~J.}\
  \bibnamefont {Herrmann}},\ }\href@noop {} {\bibfield  {journal} {\bibinfo
  {journal} {P. Natl. Acad. Sci. USA},\ }\textbf {\bibinfo {volume} {104}},\
  \bibinfo {pages} {16804} (\bibinfo {year} {2007})}\BibitemShut {NoStop}%
\bibitem [{\citenamefont {Seybold}\ \emph {et~al.}(2010)\citenamefont
  {Seybold}, \citenamefont {Molnar}, \citenamefont {Akca}, \citenamefont
  {Doumi}, \citenamefont {Tavares}, \citenamefont {Shinbrot}, \citenamefont
  {Andrade}, \citenamefont {Kinzelbach},\ and\ \citenamefont
  {Herrmann}}]{Seybold10}%
  \BibitemOpen
  \bibfield  {author} {\bibinfo {author} {\bibfnamefont {H.~J.}\ \bibnamefont
  {Seybold}}, \bibinfo {author} {\bibfnamefont {P.}~\bibnamefont {Molnar}},
  \bibinfo {author} {\bibfnamefont {D.}~\bibnamefont {Akca}}, \bibinfo {author}
  {\bibfnamefont {M.}~\bibnamefont {Doumi}}, \bibinfo {author} {\bibfnamefont
  {M.~C.}\ \bibnamefont {Tavares}}, \bibinfo {author} {\bibfnamefont
  {T.}~\bibnamefont {Shinbrot}}, \bibinfo {author} {\bibfnamefont {J.~S.}\
  \bibnamefont {Andrade}}, \bibinfo {author} {\bibfnamefont {W.}~\bibnamefont
  {Kinzelbach}}, \ and\ \bibinfo {author} {\bibfnamefont {H.~J.}\ \bibnamefont
  {Herrmann}},\ }\href@noop {} {\bibfield  {journal} {\bibinfo  {journal}
  {Geophys. Res. Lett.},\ }\textbf {\bibinfo {volume} {37}},\ \bibinfo {pages}
  {L08402} (\bibinfo {year} {2010})}\BibitemShut {NoStop}%
\bibitem [{\citenamefont {Baldassarri}\ \emph {et~al.}(2008)\citenamefont
  {Baldassarri}, \citenamefont {Montuori}, \citenamefont {Prieto-Ballesteros},\
  and\ \citenamefont {Manrubia}}]{Baldassarri08}%
  \BibitemOpen
  \bibfield  {author} {\bibinfo {author} {\bibfnamefont {A.}~\bibnamefont
  {Baldassarri}}, \bibinfo {author} {\bibfnamefont {M.}~\bibnamefont
  {Montuori}}, \bibinfo {author} {\bibfnamefont {O.}~\bibnamefont
  {Prieto-Ballesteros}}, \ and\ \bibinfo {author} {\bibfnamefont {S.~C.}\
  \bibnamefont {Manrubia}},\ }\href@noop {} {\bibfield  {journal} {\bibinfo
  {journal} {J. Geophys. Res.},\ }\textbf {\bibinfo {volume} {113}},\ \bibinfo
  {pages} {E09002} (\bibinfo {year} {2008})}\BibitemShut {NoStop}%
\bibitem [{\citenamefont {Boffetta}\ \emph {et~al.}(2008)\citenamefont
  {Boffetta}, \citenamefont {Celani}, \citenamefont {Dezzani},\ and\
  \citenamefont {Seminara}}]{Boffetta08}%
  \BibitemOpen
  \bibfield  {author} {\bibinfo {author} {\bibfnamefont {G.}~\bibnamefont
  {Boffetta}}, \bibinfo {author} {\bibfnamefont {A.}~\bibnamefont {Celani}},
  \bibinfo {author} {\bibfnamefont {D.}~\bibnamefont {Dezzani}}, \ and\
  \bibinfo {author} {\bibfnamefont {A.}~\bibnamefont {Seminara}},\ }\href@noop
  {} {\bibfield  {journal} {\bibinfo  {journal} {Geophys. Res. Lett.},\
  }\textbf {\bibinfo {volume} {35}},\ \bibinfo {pages} {L03615} (\bibinfo
  {year} {2008})}\BibitemShut {NoStop}%
\bibitem [{\citenamefont {Fehr}\ \emph {et~al.}(2009)\citenamefont {Fehr},
  \citenamefont {\mbox{Andrade Jr.}}, \citenamefont {\mbox{da Cunha}},
  \citenamefont {\mbox{da Silva}}, \citenamefont {Herrmann}, \citenamefont
  {Kadau}, \citenamefont {Moukarzel},\ and\ \citenamefont {Oliveira}}]{Fehr09}%
  \BibitemOpen
  \bibfield  {author} {\bibinfo {author} {\bibfnamefont {E.}~\bibnamefont
  {Fehr}}, \bibinfo {author} {\bibfnamefont {J.~S.}\ \bibnamefont
  {\mbox{Andrade Jr.}}}, \bibinfo {author} {\bibfnamefont {S.~D.}\ \bibnamefont
  {\mbox{da Cunha}}}, \bibinfo {author} {\bibfnamefont {L.~R.}\ \bibnamefont
  {\mbox{da Silva}}}, \bibinfo {author} {\bibfnamefont {H.~J.}\ \bibnamefont
  {Herrmann}}, \bibinfo {author} {\bibfnamefont {D.}~\bibnamefont {Kadau}},
  \bibinfo {author} {\bibfnamefont {C.~F.}\ \bibnamefont {Moukarzel}}, \ and\
  \bibinfo {author} {\bibfnamefont {E.~A.}\ \bibnamefont {Oliveira}},\
  }\href@noop {} {\bibfield  {journal} {\bibinfo  {journal} {J. Stat. Mech.},\
  \bibinfo {pages} {P09007}} (\bibinfo {year} {2009})}\BibitemShut {NoStop}%
\bibitem [{\citenamefont {Fehr}\ \emph {et~al.}(2011)\citenamefont {Fehr},
  \citenamefont {Kadau}, \citenamefont {\mbox{Andrade Jr.}},\ and\
  \citenamefont {Herrmann}}]{Fehr11}%
  \BibitemOpen
  \bibfield  {author} {\bibinfo {author} {\bibfnamefont {E.}~\bibnamefont
  {Fehr}}, \bibinfo {author} {\bibfnamefont {D.}~\bibnamefont {Kadau}},
  \bibinfo {author} {\bibfnamefont {J.~S.}\ \bibnamefont {\mbox{Andrade Jr.}}},
  \ and\ \bibinfo {author} {\bibfnamefont {H.~J.}\ \bibnamefont {Herrmann}},\
  }\href@noop {} {\bibfield  {journal} {\bibinfo  {journal} {Phys. Rev.
  Lett.},\ }\textbf {\bibinfo {volume} {106}},\ \bibinfo {pages} {048501}
  (\bibinfo {year} {2011})}\BibitemShut {NoStop}%
\bibitem [{\citenamefont {\mbox{Andrade Jr.}}\ \emph
  {et~al.}(2010)\citenamefont {\mbox{Andrade Jr.}}, \citenamefont {Reis},
  \citenamefont {Oliveira}, \citenamefont {Fehr},\ and\ \citenamefont
  {Herrmann}}]{Andrade11}%
  \BibitemOpen
  \bibfield  {author} {\bibinfo {author} {\bibfnamefont {J.~S.}\ \bibnamefont
  {\mbox{Andrade Jr.}}}, \bibinfo {author} {\bibfnamefont {S.~D.~S.}\
  \bibnamefont {Reis}}, \bibinfo {author} {\bibfnamefont {E.~A.}\ \bibnamefont
  {Oliveira}}, \bibinfo {author} {\bibfnamefont {E.}~\bibnamefont {Fehr}}, \
  and\ \bibinfo {author} {\bibfnamefont {H.~J.}\ \bibnamefont {Herrmann}},\
  }\href@noop {} {\bibfield  {journal} {\bibinfo  {journal} {Comput. Sci.
  Eng.},\ }\textbf {\bibinfo {volume} {13}},\ \bibinfo {pages} {74} (\bibinfo
  {year} {2010})}\BibitemShut {NoStop}%
\bibitem [{\citenamefont {Vorosmarty}\ \emph {et~al.}(1998)\citenamefont
  {Vorosmarty}, \citenamefont {Federer},\ and\ \citenamefont
  {Schloss}}]{Vorosmarty98}%
  \BibitemOpen
  \bibfield  {author} {\bibinfo {author} {\bibfnamefont {C.~J.}\ \bibnamefont
  {Vorosmarty}}, \bibinfo {author} {\bibfnamefont {C.~A.}\ \bibnamefont
  {Federer}}, \ and\ \bibinfo {author} {\bibfnamefont {A.~L.}\ \bibnamefont
  {Schloss}},\ }\href@noop {} {\bibfield  {journal} {\bibinfo  {journal} {J.
  Hydrol.},\ }\textbf {\bibinfo {volume} {207}},\ \bibinfo {pages} {147}
  (\bibinfo {year} {1998})}\BibitemShut {NoStop}%
\bibitem [{\citenamefont {Kwarteng}\ \emph {et~al.}(2000)\citenamefont
  {Kwarteng}, \citenamefont {Viswanathan}, \citenamefont {Al-Senafy},\ and\
  \citenamefont {Rashid}}]{Kwarteng00}%
  \BibitemOpen
  \bibfield  {author} {\bibinfo {author} {\bibfnamefont {A.~Y.}\ \bibnamefont
  {Kwarteng}}, \bibinfo {author} {\bibfnamefont {M.~N.}\ \bibnamefont
  {Viswanathan}}, \bibinfo {author} {\bibfnamefont {M.~N.}\ \bibnamefont
  {Al-Senafy}}, \ and\ \bibinfo {author} {\bibfnamefont {T.}~\bibnamefont
  {Rashid}},\ }\href@noop {} {\bibfield  {journal} {\bibinfo  {journal} {J.
  Arid. Environ.},\ }\textbf {\bibinfo {volume} {46}},\ \bibinfo {pages} {137}
  (\bibinfo {year} {2000})}\BibitemShut {NoStop}%
\bibitem [{\citenamefont {Sarangi}\ and\ \citenamefont
  {Bhattacharya}(2005)}]{Sarangi05}%
  \BibitemOpen
  \bibfield  {author} {\bibinfo {author} {\bibfnamefont {A.}~\bibnamefont
  {Sarangi}}\ and\ \bibinfo {author} {\bibfnamefont {A.~K.}\ \bibnamefont
  {Bhattacharya}},\ }\href@noop {} {\bibfield  {journal} {\bibinfo  {journal}
  {Agric. Water Manage.},\ }\textbf {\bibinfo {volume} {78}},\ \bibinfo {pages}
  {195} (\bibinfo {year} {2005})}\BibitemShut {NoStop}%
\bibitem [{\citenamefont {Dhakal}\ and\ \citenamefont
  {Sidle}(2004)}]{Dhakal04}%
  \BibitemOpen
  \bibfield  {author} {\bibinfo {author} {\bibfnamefont {A.~S.}\ \bibnamefont
  {Dhakal}}\ and\ \bibinfo {author} {\bibfnamefont {R.~C.}\ \bibnamefont
  {Sidle}},\ }\href@noop {} {\bibfield  {journal} {\bibinfo  {journal} {Hydrol.
  Processes},\ }\textbf {\bibinfo {volume} {18}},\ \bibinfo {pages} {757}
  (\bibinfo {year} {2004})}\BibitemShut {NoStop}%
\bibitem [{\citenamefont {Pradhan}\ \emph {et~al.}(2006)\citenamefont
  {Pradhan}, \citenamefont {Singh},\ and\ \citenamefont
  {Buchroithner}}]{Pradhan06}%
  \BibitemOpen
  \bibfield  {author} {\bibinfo {author} {\bibfnamefont {B.}~\bibnamefont
  {Pradhan}}, \bibinfo {author} {\bibfnamefont {R.~P.}\ \bibnamefont {Singh}},
  \ and\ \bibinfo {author} {\bibfnamefont {M.~F.}\ \bibnamefont
  {Buchroithner}},\ }\href@noop {} {\bibfield  {journal} {\bibinfo  {journal}
  {Adv. Space Res.},\ }\textbf {\bibinfo {volume} {37}},\ \bibinfo {pages}
  {698} (\bibinfo {year} {2006})}\BibitemShut {NoStop}%
\bibitem [{\citenamefont {Lazzari}\ \emph {et~al.}(2006)\citenamefont
  {Lazzari}, \citenamefont {Geraldi}, \citenamefont {Lapenna},\ and\
  \citenamefont {Loperte}}]{Lazzari06}%
  \BibitemOpen
  \bibfield  {author} {\bibinfo {author} {\bibfnamefont {M.}~\bibnamefont
  {Lazzari}}, \bibinfo {author} {\bibfnamefont {E.}~\bibnamefont {Geraldi}},
  \bibinfo {author} {\bibfnamefont {V.}~\bibnamefont {Lapenna}}, \ and\
  \bibinfo {author} {\bibfnamefont {A.}~\bibnamefont {Loperte}},\ }\href@noop
  {} {\bibfield  {journal} {\bibinfo  {journal} {Landslides},\ }\textbf
  {\bibinfo {volume} {3}},\ \bibinfo {pages} {275} (\bibinfo {year}
  {2006})}\BibitemShut {NoStop}%
\bibitem [{\citenamefont {Lee}\ and\ \citenamefont {Lin}(2006)}]{Lee06}%
  \BibitemOpen
  \bibfield  {author} {\bibinfo {author} {\bibfnamefont {K.~T.}\ \bibnamefont
  {Lee}}\ and\ \bibinfo {author} {\bibfnamefont {Y.~T.}\ \bibnamefont {Lin}},\
  }\href@noop {} {\bibfield  {journal} {\bibinfo  {journal} {J. Am. Water
  Resour. Assoc.},\ }\textbf {\bibinfo {volume} {42}},\ \bibinfo {pages} {1615}
  (\bibinfo {year} {2006})}\BibitemShut {NoStop}%
\bibitem [{\citenamefont {Burlando}\ \emph {et~al.}(1994)\citenamefont
  {Burlando}, \citenamefont {Mancini},\ and\ \citenamefont
  {Rosso}}]{Burlando94}%
  \BibitemOpen
  \bibfield  {author} {\bibinfo {author} {\bibfnamefont {P.}~\bibnamefont
  {Burlando}}, \bibinfo {author} {\bibfnamefont {M.}~\bibnamefont {Mancini}}, \
  and\ \bibinfo {author} {\bibfnamefont {R.}~\bibnamefont {Rosso}},\
  }\href@noop {} {\bibfield  {journal} {\bibinfo  {journal} {IFIP Trans. B},\
  }\textbf {\bibinfo {volume} {16}},\ \bibinfo {pages} {91} (\bibinfo {year}
  {1994})}\BibitemShut {NoStop}%
\bibitem [{\citenamefont {Yang}\ \emph {et~al.}(2007)\citenamefont {Yang},
  \citenamefont {Zhao}, \citenamefont {Armstrong}, \citenamefont {Robinson},\
  and\ \citenamefont {Brodzik}}]{Yang07}%
  \BibitemOpen
  \bibfield  {author} {\bibinfo {author} {\bibfnamefont {D.~Q.}\ \bibnamefont
  {Yang}}, \bibinfo {author} {\bibfnamefont {Y.}~\bibnamefont {Zhao}}, \bibinfo
  {author} {\bibfnamefont {R.}~\bibnamefont {Armstrong}}, \bibinfo {author}
  {\bibfnamefont {D.}~\bibnamefont {Robinson}}, \ and\ \bibinfo {author}
  {\bibfnamefont {M.}~\bibnamefont {Brodzik}},\ }\href@noop {} {\bibfield
  {journal} {\bibinfo  {journal} {J. Geophys. Res. Earth Surf.},\ }\textbf
  {\bibinfo {volume} {112}},\ \bibinfo {pages} {F02S22} (\bibinfo {year}
  {2007})}\BibitemShut {NoStop}%
\bibitem [{\citenamefont {Penning-Roswell}\ \emph {et~al.}(1992)\citenamefont
  {Penning-Roswell}, \citenamefont {Green}, \citenamefont {Thompson},
  \citenamefont {Coker}, \citenamefont {Tunstall}, \citenamefont {Richards},\
  and\ \citenamefont {Parker}}]{Penning-Roswell92}%
  \BibitemOpen
  \bibfield  {author} {\bibinfo {author} {\bibfnamefont {E.~C.}\ \bibnamefont
  {Penning-Roswell}}, \bibinfo {author} {\bibfnamefont {C.~H.}\ \bibnamefont
  {Green}}, \bibinfo {author} {\bibfnamefont {P.~M.}\ \bibnamefont {Thompson}},
  \bibinfo {author} {\bibfnamefont {A.~M.}\ \bibnamefont {Coker}}, \bibinfo
  {author} {\bibfnamefont {S.~M.}\ \bibnamefont {Tunstall}}, \bibinfo {author}
  {\bibfnamefont {C.}~\bibnamefont {Richards}}, \ and\ \bibinfo {author}
  {\bibfnamefont {D.~J.}\ \bibnamefont {Parker}},\ }\href@noop {} {\emph
  {\bibinfo {title} {The Economics of Coastal Management}}}\ (\bibinfo
  {publisher} {Belhaven Press},\ \bibinfo {address} {London},\ \bibinfo {year}
  {1992})\BibitemShut {NoStop}%
\bibitem [{\citenamefont {Sapoval}\ \emph {et~al.}(2004)\citenamefont
  {Sapoval}, \citenamefont {Baldassarri},\ and\ \citenamefont
  {Gabrielli}}]{Sapoval04}%
  \BibitemOpen
  \bibfield  {author} {\bibinfo {author} {\bibfnamefont {B.}~\bibnamefont
  {Sapoval}}, \bibinfo {author} {\bibfnamefont {A.}~\bibnamefont
  {Baldassarri}}, \ and\ \bibinfo {author} {\bibfnamefont {A.}~\bibnamefont
  {Gabrielli}},\ }\href@noop {} {\bibfield  {journal} {\bibinfo  {journal}
  {Phys. Rev. Lett.},\ }\textbf {\bibinfo {volume} {93}},\ \bibinfo {pages}
  {098501} (\bibinfo {year} {2004})}\BibitemShut {NoStop}%
\bibitem [{\citenamefont {Bak}(1996)}]{Bak96}%
  \BibitemOpen
  \bibfield  {author} {\bibinfo {author} {\bibfnamefont {P.}~\bibnamefont
  {Bak}},\ }\href@noop {} {\emph {\bibinfo {title} {How nature works? The
  science of self-organized criticality}}}\ (\bibinfo  {publisher}
  {Springer-Verlag New York},\ \bibinfo {address} {United States of America},\
  \bibinfo {year} {1996})\BibitemShut {NoStop}%
\bibitem [{\citenamefont {Turcotte}(1999)}]{Turcotte99}%
  \BibitemOpen
  \bibfield  {author} {\bibinfo {author} {\bibfnamefont {D.~L.}\ \bibnamefont
  {Turcotte}},\ }\href@noop {} {\bibfield  {journal} {\bibinfo  {journal} {Rep.
  Prog. Phys.},\ }\textbf {\bibinfo {volume} {62}},\ \bibinfo {pages} {1377}
  (\bibinfo {year} {1999})}\BibitemShut {NoStop}%
\bibitem [{\citenamefont {Sapoval}\ \emph {et~al.}(1985)\citenamefont
  {Sapoval}, \citenamefont {Rosso},\ and\ \citenamefont {Gouyet}}]{Sapoval85}%
  \BibitemOpen
  \bibfield  {author} {\bibinfo {author} {\bibfnamefont {B.}~\bibnamefont
  {Sapoval}}, \bibinfo {author} {\bibfnamefont {M.}~\bibnamefont {Rosso}}, \
  and\ \bibinfo {author} {\bibfnamefont {J.~F.}\ \bibnamefont {Gouyet}},\
  }\href@noop {} {\bibfield  {journal} {\bibinfo  {journal} {J. Physique
  Lett.},\ }\textbf {\bibinfo {volume} {46}},\ \bibinfo {pages} {L149}
  (\bibinfo {year} {1985})}\BibitemShut {NoStop}%
\bibitem [{\citenamefont {Grossman}\ and\ \citenamefont
  {Aharony}(1987)}]{Grossman87}%
  \BibitemOpen
  \bibfield  {author} {\bibinfo {author} {\bibfnamefont {T.}~\bibnamefont
  {Grossman}}\ and\ \bibinfo {author} {\bibfnamefont {A.}~\bibnamefont
  {Aharony}},\ }\href@noop {} {\bibfield  {journal} {\bibinfo  {journal} {J.
  Phys. A},\ }\textbf {\bibinfo {volume} {20}},\ \bibinfo {pages} {L1193}
  (\bibinfo {year} {1987})}\BibitemShut {NoStop}%
\bibitem [{\citenamefont {Richardson}(1961)}]{Richardson61}%
  \BibitemOpen
  \bibfield  {author} {\bibinfo {author} {\bibfnamefont {L.~F.}\ \bibnamefont
  {Richardson}},\ }\href@noop {} {\bibfield  {journal} {\bibinfo  {journal}
  {Gen. Syst. Yearb.},\ }\textbf {\bibinfo {volume} {6}},\ \bibinfo {pages}
  {139} (\bibinfo {year} {1961})}\BibitemShut {NoStop}%
\bibitem [{\citenamefont {Peitgen}\ and\ \citenamefont
  {Saupe}(1988)}]{Peitgen88}%
  \BibitemOpen
  \bibinfo {editor} {\bibfnamefont {H.}~\bibnamefont {Peitgen}}\ and\ \bibinfo
  {editor} {\bibfnamefont {D.}~\bibnamefont {Saupe}},\ eds.,\ \href@noop {}
  {\emph {\bibinfo {title} {The Science of Fractal Images}}}\ (\bibinfo
  {publisher} {Springer},\ \bibinfo {address} {New York},\ \bibinfo {year}
  {1988})\BibitemShut {NoStop}%
\bibitem [{\citenamefont {Sahimi}(1994)}]{Sahimi94}%
  \BibitemOpen
  \bibfield  {author} {\bibinfo {author} {\bibfnamefont {M.}~\bibnamefont
  {Sahimi}},\ }\href@noop {} {\bibfield  {journal} {\bibinfo  {journal} {J.
  Phys. I France},\ }\textbf {\bibinfo {volume} {4}},\ \bibinfo {pages} {1263}
  (\bibinfo {year} {1994})}\BibitemShut {NoStop}%
\bibitem [{\citenamefont {Sahimi}\ and\ \citenamefont
  {Mukhopadhyay}(1996)}]{Sahimi96}%
  \BibitemOpen
  \bibfield  {author} {\bibinfo {author} {\bibfnamefont {M.}~\bibnamefont
  {Sahimi}}\ and\ \bibinfo {author} {\bibfnamefont {S.}~\bibnamefont
  {Mukhopadhyay}},\ }\href@noop {} {\bibfield  {journal} {\bibinfo  {journal}
  {Phys. Rev. E},\ }\textbf {\bibinfo {volume} {54}},\ \bibinfo {pages} {3870}
  (\bibinfo {year} {1996})}\BibitemShut {NoStop}%
\bibitem [{\citenamefont {Makse}\ \emph {et~al.}(1996)\citenamefont {Makse},
  \citenamefont {Havlin}, \citenamefont {Schwartz},\ and\ \citenamefont
  {Stanley}}]{Makse96}%
  \BibitemOpen
  \bibfield  {author} {\bibinfo {author} {\bibfnamefont {H.~A.}\ \bibnamefont
  {Makse}}, \bibinfo {author} {\bibfnamefont {S.}~\bibnamefont {Havlin}},
  \bibinfo {author} {\bibfnamefont {M.}~\bibnamefont {Schwartz}}, \ and\
  \bibinfo {author} {\bibfnamefont {H.~E.}\ \bibnamefont {Stanley}},\
  }\href@noop {} {\bibfield  {journal} {\bibinfo  {journal} {Phys. Rev. E},\
  }\textbf {\bibinfo {volume} {53}},\ \bibinfo {pages} {5445} (\bibinfo {year}
  {1996})}\BibitemShut {NoStop}%
\bibitem [{\citenamefont {Oliveira}\ \emph {et~al.}(2011)\citenamefont
  {Oliveira}, \citenamefont {Schrenk}, \citenamefont {Ara\'ujo}, \citenamefont
  {Herrmann},\ and\ \citenamefont {\mbox{Andrade Jr.}}}]{Oliveira11}%
  \BibitemOpen
  \bibfield  {author} {\bibinfo {author} {\bibfnamefont {E.~A.}\ \bibnamefont
  {Oliveira}}, \bibinfo {author} {\bibfnamefont {K.~J.}\ \bibnamefont
  {Schrenk}}, \bibinfo {author} {\bibfnamefont {N.~A.~M.}\ \bibnamefont
  {Ara\'ujo}}, \bibinfo {author} {\bibfnamefont {H.~J.}\ \bibnamefont
  {Herrmann}}, \ and\ \bibinfo {author} {\bibfnamefont {J.~S.}\ \bibnamefont
  {\mbox{Andrade Jr.}}},\ }\href@noop {} {\bibfield  {journal} {\bibinfo
  {journal} {Phys. Rev. E},\ }\textbf {\bibinfo {volume} {83}},\ \bibinfo
  {pages} {046113} (\bibinfo {year} {2011})}\BibitemShut {NoStop}%
\bibitem [{\citenamefont {Kardar}\ \emph {et~al.}(1986)\citenamefont {Kardar},
  \citenamefont {Parisi},\ and\ \citenamefont {Zhang}}]{Kardar86}%
  \BibitemOpen
  \bibfield  {author} {\bibinfo {author} {\bibfnamefont {M.}~\bibnamefont
  {Kardar}}, \bibinfo {author} {\bibfnamefont {G.}~\bibnamefont {Parisi}}, \
  and\ \bibinfo {author} {\bibfnamefont {Y.~C.}\ \bibnamefont {Zhang}},\
  }\href@noop {} {\bibfield  {journal} {\bibinfo  {journal} {Phys. Rev.
  Lett.},\ }\textbf {\bibinfo {volume} {56}},\ \bibinfo {pages} {889} (\bibinfo
  {year} {1986})}\BibitemShut {NoStop}%
\bibitem [{\citenamefont {Tricot}\ \emph {et~al.}(1988)\citenamefont {Tricot},
  \citenamefont {Quiniou}, \citenamefont {Wehbi}, \citenamefont
  {Roquescarmes},\ and\ \citenamefont {Dubuc}}]{Tricot88}%
  \BibitemOpen
  \bibfield  {author} {\bibinfo {author} {\bibfnamefont {C.}~\bibnamefont
  {Tricot}}, \bibinfo {author} {\bibfnamefont {J.~F.}\ \bibnamefont {Quiniou}},
  \bibinfo {author} {\bibfnamefont {D.}~\bibnamefont {Wehbi}}, \bibinfo
  {author} {\bibfnamefont {C.}~\bibnamefont {Roquescarmes}}, \ and\ \bibinfo
  {author} {\bibfnamefont {B.}~\bibnamefont {Dubuc}},\ }\href@noop {}
  {\bibfield  {journal} {\bibinfo  {journal} {Revue Phys. Appl.},\ }\textbf
  {\bibinfo {volume} {23}},\ \bibinfo {pages} {111} (\bibinfo {year}
  {1988})}\BibitemShut {NoStop}%
\bibitem [{\citenamefont {Barab\'asi}\ and\ \citenamefont
  {Stanley}(1995)}]{Barabasi95}%
  \BibitemOpen
  \bibfield  {author} {\bibinfo {author} {\bibfnamefont {A.-L.}\ \bibnamefont
  {Barab\'asi}}\ and\ \bibinfo {author} {\bibfnamefont {H.~E.}\ \bibnamefont
  {Stanley}},\ }\href@noop {} {\emph {\bibinfo {title} {Fractal Concepts in
  Surface Growth}}}\ (\bibinfo  {publisher} {Cambridge University Press},\
  \bibinfo {address} {United Kingdom},\ \bibinfo {year} {1995})\BibitemShut
  {NoStop}%
\bibitem [{\citenamefont {Family}\ and\ \citenamefont
  {Vicsek}(1985)}]{Family85}%
  \BibitemOpen
  \bibfield  {author} {\bibinfo {author} {\bibfnamefont {F.}~\bibnamefont
  {Family}}\ and\ \bibinfo {author} {\bibfnamefont {T.}~\bibnamefont
  {Vicsek}},\ }\href@noop {} {\bibfield  {journal} {\bibinfo  {journal} {J.
  Phys. A},\ }\textbf {\bibinfo {volume} {18}},\ \bibinfo {pages} {L75}
  (\bibinfo {year} {1985})}\BibitemShut {NoStop}%
\bibitem [{\citenamefont {\'Odor}(2004)}]{Odor04}%
  \BibitemOpen
  \bibfield  {author} {\bibinfo {author} {\bibfnamefont {G.}~\bibnamefont
  {\'Odor}},\ }\href@noop {} {\bibfield  {journal} {\bibinfo  {journal} {Rev.
  Mod. Phys.},\ }\textbf {\bibinfo {volume} {76}},\ \bibinfo {pages} {663}
  (\bibinfo {year} {2004})}\BibitemShut {NoStop}%
\bibitem [{\citenamefont {Prakash}\ \emph {et~al.}(1992)\citenamefont
  {Prakash}, \citenamefont {Havlin}, \citenamefont {Schwartz},\ and\
  \citenamefont {Stanley}}]{Prakash92}%
  \BibitemOpen
  \bibfield  {author} {\bibinfo {author} {\bibfnamefont {S.}~\bibnamefont
  {Prakash}}, \bibinfo {author} {\bibfnamefont {S.}~\bibnamefont {Havlin}},
  \bibinfo {author} {\bibfnamefont {M.}~\bibnamefont {Schwartz}}, \ and\
  \bibinfo {author} {\bibfnamefont {H.~E.}\ \bibnamefont {Stanley}},\
  }\href@noop {} {\bibfield  {journal} {\bibinfo  {journal} {Phys. Rev. A},\
  }\textbf {\bibinfo {volume} {46}},\ \bibinfo {pages} {R1724} (\bibinfo {year}
  {1992})}\BibitemShut {NoStop}%
\bibitem [{\citenamefont {Kikkinides}\ and\ \citenamefont
  {Burganos}(1999)}]{Kikkinides99}%
  \BibitemOpen
  \bibfield  {author} {\bibinfo {author} {\bibfnamefont {E.~S.}\ \bibnamefont
  {Kikkinides}}\ and\ \bibinfo {author} {\bibfnamefont {V.~N.}\ \bibnamefont
  {Burganos}},\ }\href@noop {} {\bibfield  {journal} {\bibinfo  {journal}
  {Phys. Rev. E},\ }\textbf {\bibinfo {volume} {59}},\ \bibinfo {pages} {7185}
  (\bibinfo {year} {1999})}\BibitemShut {NoStop}%
\bibitem [{\citenamefont {Stanley}\ \emph {et~al.}(1999)\citenamefont
  {Stanley}, \citenamefont {\mbox{Andrade Jr.}}, \citenamefont {Havlin},
  \citenamefont {Makse},\ and\ \citenamefont {Suki}}]{Stanley99}%
  \BibitemOpen
  \bibfield  {author} {\bibinfo {author} {\bibfnamefont {H.~E.}\ \bibnamefont
  {Stanley}}, \bibinfo {author} {\bibfnamefont {J.~S.}\ \bibnamefont
  {\mbox{Andrade Jr.}}}, \bibinfo {author} {\bibfnamefont {S.}~\bibnamefont
  {Havlin}}, \bibinfo {author} {\bibfnamefont {H.~A.}\ \bibnamefont {Makse}}, \
  and\ \bibinfo {author} {\bibfnamefont {B.}~\bibnamefont {Suki}},\ }\href@noop
  {} {\bibfield  {journal} {\bibinfo  {journal} {Physica A},\ }\textbf
  {\bibinfo {volume} {266}},\ \bibinfo {pages} {5} (\bibinfo {year}
  {1999})}\BibitemShut {NoStop}%
\bibitem [{\citenamefont {Makse}\ \emph {et~al.}(2000)\citenamefont {Makse},
  \citenamefont {\mbox{Andrade Jr.}},\ and\ \citenamefont {Stanley}}]{Makse00}%
  \BibitemOpen
  \bibfield  {author} {\bibinfo {author} {\bibfnamefont {H.~A.}\ \bibnamefont
  {Makse}}, \bibinfo {author} {\bibfnamefont {J.~S.}\ \bibnamefont
  {\mbox{Andrade Jr.}}}, \ and\ \bibinfo {author} {\bibfnamefont {H.~E.}\
  \bibnamefont {Stanley}},\ }\href@noop {} {\bibfield  {journal} {\bibinfo
  {journal} {Phys. Rev. E},\ }\textbf {\bibinfo {volume} {61}},\ \bibinfo
  {pages} {583} (\bibinfo {year} {2000})}\BibitemShut {NoStop}%
\bibitem [{\citenamefont {Ara\'ujo}\ \emph {et~al.}(2002)\citenamefont
  {Ara\'ujo}, \citenamefont {Moreira}, \citenamefont {Makse}, \citenamefont
  {Stanley},\ and\ \citenamefont {\mbox{Andrade Jr.}}}]{Araujo02}%
  \BibitemOpen
  \bibfield  {author} {\bibinfo {author} {\bibfnamefont {A.~D.}\ \bibnamefont
  {Ara\'ujo}}, \bibinfo {author} {\bibfnamefont {A.~A.}\ \bibnamefont
  {Moreira}}, \bibinfo {author} {\bibfnamefont {H.~A.}\ \bibnamefont {Makse}},
  \bibinfo {author} {\bibfnamefont {H.~E.}\ \bibnamefont {Stanley}}, \ and\
  \bibinfo {author} {\bibfnamefont {J.~S.}\ \bibnamefont {\mbox{Andrade
  Jr.}}},\ }\href@noop {} {\bibfield  {journal} {\bibinfo  {journal} {Phys.
  Rev. E},\ }\textbf {\bibinfo {volume} {66}},\ \bibinfo {pages} {046304}
  (\bibinfo {year} {2002})}\BibitemShut {NoStop}%
\bibitem [{\citenamefont {Ara\'ujo}\ \emph {et~al.}(2003)\citenamefont
  {Ara\'ujo}, \citenamefont {Moreira}, \citenamefont {\mbox{Costa Filho}},\
  and\ \citenamefont {\mbox{Andrade Jr.}}}]{Araujo03}%
  \BibitemOpen
  \bibfield  {author} {\bibinfo {author} {\bibfnamefont {A.~D.}\ \bibnamefont
  {Ara\'ujo}}, \bibinfo {author} {\bibfnamefont {A.~A.}\ \bibnamefont
  {Moreira}}, \bibinfo {author} {\bibfnamefont {R.~N.}\ \bibnamefont
  {\mbox{Costa Filho}}}, \ and\ \bibinfo {author} {\bibfnamefont {J.~S.}\
  \bibnamefont {\mbox{Andrade Jr.}}},\ }\href@noop {} {\bibfield  {journal}
  {\bibinfo  {journal} {Phys. Rev. E},\ }\textbf {\bibinfo {volume} {67}},\
  \bibinfo {pages} {027102} (\bibinfo {year} {2003})}\BibitemShut {NoStop}%
\bibitem [{\citenamefont {Du}\ \emph {et~al.}(1996)\citenamefont {Du},
  \citenamefont {Satik},\ and\ \citenamefont {Yortsos}}]{Du04}%
  \BibitemOpen
  \bibfield  {author} {\bibinfo {author} {\bibfnamefont {C.}~\bibnamefont
  {Du}}, \bibinfo {author} {\bibfnamefont {C.}~\bibnamefont {Satik}}, \ and\
  \bibinfo {author} {\bibfnamefont {Y.~C.}\ \bibnamefont {Yortsos}},\
  }\href@noop {} {\bibfield  {journal} {\bibinfo  {journal} {AIChE Journal},\
  }\textbf {\bibinfo {volume} {42}},\ \bibinfo {pages} {2392} (\bibinfo {year}
  {1996})}\BibitemShut {NoStop}%
\end{thebibliography}%

\end{document}